# Study of radiation damage induced by 12 keV X-rays in MOS structures built on high resistivity n-type silicon

Jiaguo Zhang,[a] Ioana Pintilie,[b,a*] Eckhart Fretwurst,[a] Robert Klanner,[a] Hanno Perrey[a] and Joern Schwandt[a]

[a]*Institute for Experimental Physics, University of Hamburg, Germany, and* [b]*National Institute of Materials Physics, Romania. E-mail: ioana@infim.ro,*

**Abstract:** Imaging experiments at the European X-ray Free Electron Laser (XFEL) require silicon pixel sensors with extraordinary performance specifications: Doses of up to 1 GGy of 12 keV photons, up to $10^5$ 12 keV photons per pixel of 200 µm × 200 µm arriving within less than 100 fs, and a time interval between XFEL pulses of 220 ns. To address these challenges, in particular the question of radiation damage, the properties of the $SiO_2$ layer and of the $Si$-$SiO_2$ interface using MOS capacitors manufactured on high resistivity n-type silicon irradiated to X-ray doses between 10 kGy and 1 GGy, have been studied. Measurements of Capacitance/Conductance-Voltage (C/G-V) at different frequencies, as well as of Thermal Dielectric Relaxation Current (TDRC) have been performed. The data can be described by a radiation dependent oxide charge density and three dominant radiation-induced interface states with Gaussian-like energy distributions in the silicon band gap. It is found that the densities of the fixed oxide charges and of the three interface states increase up to dose values of approximately 10 MGy and then saturate or even decrease. The shapes and the frequency dependences of the C/G-V measurements can be quantitatively described by a simple model using the parameters extracted from the TDRC measurements.

## 1. Introduction

The European X-ray Free Electron Laser (XFEL), planned to be operational in 2014, will provide fully coherent, shorter than 100 fs long X-ray pulses, each with up to $10^{12}$ 12 keV photons. For its science potential we refer to Chapman *et al.*, (2007); Murnane *et al.*, (2009) and the XFEL webpage. The high brilliance of the XFEL poses unprecedented requirements for the detection systems: A dynamic range from single 12 keV photons to $10^5$ photons deposited in less than 100 fs, a time between XFEL pulses of 220 ns, and radiation doses up to 1 GGy for 3 years of operation (Graafsma, 2009). To optimize sensors for these requirements demands a good understanding of the radiation damage caused by X-rays.

In this work, MOS capacitors manufactured on high resistivity n-type silicon were used to experimentally determine the parameters needed to understand and to simulate the performance of segmented silicon sensors as function of the X-ray dose: These are in particular the densities and properties of the Si-SiO$_2$ interface states and the oxide charge density $N_{ox}$. Both significantly influence the performance of segmented silicon sensors, like dark current, breakdown voltage as well as the formation of an accumulation layer below the SiO$_2$, which may cause changes in the inter-pixel capacitance, the inter-pixel resistance (Wuestenfeld, 2001) and the charge collection.

The maximum energy transfer to silicon atoms for 12 keV photons is 0.011 eV, which is well below the damage threshold of 21 eV for silicon (Akkerman *et al.*, 2001). Thus no bulk damage is expected for 12 keV photons. The situation is different for the insulator: The band gap of SiO$_2$ is 8.8 eV (Nicollian *et al.*, 1982), which is the energy required to break a Si-O bond. The average energy to produce an electron-hole pair is 18 eV (Ausman *et al.*, 1975; Benedetto *et al.*, 1986). Some of the charge carriers produced by the 12 keV photons recombine. Those remaining either reach the aluminum on top of the oxide or the Si-SiO$_2$ interface. Once holes come close to the Si-SiO$_2$ interface, a fraction of them will be trapped in the oxide close to the interface, and produce radiation induced positive fixed oxide charges (Nicollian *et al.*, 1982). The fraction of electron-hole pairs which escape recombination depends on the energy of the photons and on the field in the oxide during irradiation. In addition interface traps are produced, which have energy levels distributed throughout the silicon band gap and whose occupation therefore changes with gate voltage. In case of a second insulator layer, e.g. Si$_3$N$_4$, on top of the SiO$_2$ the above model has to be extended in such a way that an additional charge layer, which can be positively or negatively charged, builds up at the interface of the two insulators (Holmes-Siedle *et al.*, 2002).

## 2. Experiments and analysis techniques

### 2.1. Test structures

The MOS capacitors studied were fabricated by CiS (CiS Forschungsinstitut für Mikrosensorik und Photovoltaik GmbH, Erfurt, Germany) on 280±10 µm n-doped <100> substrates of 5-6 kΩ·cm resistivity (as determined from the CV curve of a close-by pad diode). The insulator is made of 350 nm SiO$_2$ covered by 50 nm Si$_3$N$_4$. The area of the ~1 µm thick aluminum contact is $A = 1.767 \times 10^{-2}$ cm$^2$. To check the dependence on crystal orientation, some measurements were also performed on MOS capacitors fabricated on a crystal with <111> orientation.

### 2.2. Irradiations

For the irradiations an X-ray irradiation facility has been set-up (Perrey, 2011) at DORIS, DESY. A "white" photon beam from a DORIS bending magnet (beam line F4) has been used. The maximum flux was at 12 keV, the full width of the energy spread about 10 keV. The dose rate was calibrated by the photocurrent in a silicon pad diode. The results agree to better than 20 % with calculations using the beam current, the field of the bending magnet and the geometry of the set-up. The doses quoted in this manuscript refer to the surface dose in $SiO_2$. The nominal dose rate of the irradiation set-up is 180 kGy/s. This dose rate could be reduced by a chopper by up to a factor 200. For the experiments performed in this study the dose rate was about 18 kGy/s, except for the irradiations to 12 kGy and to 1 GGy when dose rates of 1.3 kGy/s and 180 kGy/s (full dose rate), respectively were used. The test structures were mounted on an alumina substrate. The temperature during irradiation was typically ~25°C for the reduced dose rates and ~35°C for the full dose rate. It has been verified that to within 20% the results obtained did not strongly depend on the dose rate (Perrey, 2011).

## 2.3. Measurement and analysis techniques

The Capacitance-Voltage (C-V) and the Conductance-Voltage (G-V) measurements have been made at room temperature using an Agilent E4980A bridge. An A.C. voltage of 50 mV and a frequency range between 1 kHz and 1 MHz have been chosen.

For the determination of the parameters of the interface traps the Thermal Dielectric Relaxation Current (TDRC) technique (Simmons *et al.*, 1972; Mar *et al.*, 1974; Uranwala *et al.*, 1975; Mar *et al.*, 1975) was used: The MOS capacitor was cooled to 10 K biased in accumulation (0 V in our case) to fill the interface traps with electrons. Then a negative voltage to bias the structure in weak inversion was applied and the sample was heated to room temperature with a constant rate of $\beta = 0.183$ K/s while measuring the current $S(T)$ due to the emission of trapped charges.

To extract the energy distribution and the density of interface states, the procedure of Uranwala *et al.*, (1975) was used. In a MOS structure built on n-type silicon the relation between the temperature $T$ and the energy $E_t$ of a trap relative to the conduction band energy $E_c$ is given by:

$$E_t = 10^{-4}T \cdot \left[1.92 \cdot \log_{10}\left(\frac{\nu}{\beta}\right) + 3.2\right] eV/K - 0.0155\, eV \tag{1}$$

with the frequency factor $\nu = N_c \cdot v_{th} \cdot \sigma_n$. (2)

$N_c$ the density of states in the conduction band of silicon, $v_{th}$ is the average thermal velocity of the electrons and $\sigma_n$ is the capture cross section for electrons, for which a $T^{-2}$ dependence is assumed (in this case the frequency factor does not depend on temperature). The capture cross section can be determined from TDRC spectra taken at different heating rates $\beta$ using the method of Uranwala *et al.*, (1975). We will come back later to its determination and the limitations of this method for overlapping states and states with short annealing times at room temperature.

The density of interface states $D_{it}(E_t)$ is obtained from $S(T)$ according to Mar *et al.*, (1974) and Uranwala *et al.*, (1975):

$$D_{it}(E_t) = \frac{S(T)}{q \cdot A \cdot \beta \cdot 10^{-4} \cdot \left[1.92 \cdot \log_{10}\left(\frac{v}{\beta}\right) + 3.2\right] \frac{eV}{K}} \tag{3}$$

with $q$ the elementary charge, and $A$ the area of the MOS capacitor.

As discussed later, the measurements can be described by three interface states $D_{it}^j(E_t)$, $j = 1, 2, 3$. As function of the band bending $\Psi_s$, assuming that charge carriers are exchanged only with the conduction band, each state $j$ contributes to the capacitance $C$ and the conductance $G$ (Nicollian *et al.*, 1982; Pintilie *et al.*, 2010):

$$C_{it}^j(\Psi_s) = \frac{q^2}{kT} \cdot A \cdot \int_0^{E_g} D_{it}^j(E_t) \cdot \frac{f_t^0(E_t) \cdot (1 - f_t^0(E_t))}{(1 + \omega^2 \cdot \tau_j(E_t)^2)} dE_t \tag{4}$$

$$f_t^0(E_t) = \frac{1}{1 + e^{\frac{-E_t - q\Psi_s - q\eta}{kT}}} \tag{5a}$$

$$\text{with} \quad \tau_j(E_t) = \frac{1}{\sigma_n^j v_{th} N_d} \frac{1}{\left(e^{\frac{q\Psi_s}{kT}} + e^{\frac{-E_t - q\eta}{kT}}\right)} \qquad \eta = \frac{kT}{q} \cdot \ln\frac{N_d}{N_c} \tag{5b, c}$$

$$G_{it}^j(\Psi_s) = \frac{q^2 \cdot \omega^2}{kT} \cdot A \cdot \int_0^{E_g} D_{it}^j(E_t) \cdot \tau_j(E_t) \frac{f_t^0(E_t) \cdot (1 - f_t^0(E_t))}{(1 + \omega^2 \cdot \tau_j(E_t)^2)} dE_t \tag{6}$$

$\sigma_n^j$ is the electron capture cross section for state $j$, and $N_d$ the doping concentration of the silicon crystal close to the Si-SiO$_2$ interface.

The relation between the bias voltage $V_G$ applied at the gate, and the band bending $\Psi_s$ in steady state conditions (ignoring the 50 mV A.C. voltage of the capacitance bridge) is (Nicollian *et al.*, 1982; Pintilie *et al.*, 2010):

$$V_G(\Psi_s) = \Psi_s - \frac{Q_{ox}}{C_{ox}} - \frac{Q_s(\Psi_s)}{C_{ox}} - \frac{\sum_{j=1}^{3} Q_{it}^j(\Psi_s)}{C_{ox}} \tag{7}$$

where $Q_{ox}$ is the fixed charge in the isolator layer, made of $SiO_2$ and $Si_3N_4$ for the test structures, $Q_s$ the surface charge, and $Q_{it}^j$ the charge stored in the interface state $j$. Considering all interface states as acceptor-type, the charge can be expressed as (Nicollian *et al.*, 1982):

$$Q_{it}^j(\Psi_s) = -q \cdot A \cdot \int_0^{E_g} D_{it}^j(E_t) \cdot f_t^0(E_t, \Psi_s) dE_t \tag{8}$$

Whereas $Q_s$ and $Q_{it}^j$ can be calculated or determined from the TDRC measurements using the equations and references given, $Q_{ox}$ has to be obtained from a fit of the calculated C/G-V curves to the measured ones, taking into account the voltage shift due to $Q_{it}^j$. We use the equivalent circuit shown in Fig. 1 to calculate the C/G-V curves as function of the band bending $\Psi_s$ for different frequencies (Schroder *et al.*, 2000). $C_{ox}$ and $G_{leak}$ are the capacitance and conductance of the $SiO_2$-$Si_3N_4$ layer. From the fit of the C/V measurements to the model we find values of $G_{leak}$ of about $10^{-9}$ S. The value of $C_{ox}$ is obtained from the capacitance measured when the MOS capacitor is biased in accumulation. $C_d$ is the depletion capacitance of the silicon bulk below the Si-$SiO_2$ interface, which can be calculated using the Lindner approximation (Lindner, 1961; Brews, 1974). The capacitance and conductance due to interface traps are described by $C_{it}^j$ and $G_{it}^j$. The values are obtained from the TDRC measurements according to (4) and (6). $C_i$ is the inversion capacitance due to the minority carriers accumulated below the Si-$SiO_2$ interface, $G_r = 1/R_r$ is the sum of recombination and generation conductance due to the traps in the depleted silicon bulk. $C_i$ and $G_r$, which are only required to describe the C/G-V measurements in strong inversion for frequencies below a few kHz, are not relevant for the data discussed in this paper. $R_{bulk} = 1/G_{bulk}$ is the bulk resistance, which has to be taken into account because of the high resistivity of the non-depleted silicon. The frequency dependence of the conductance G in accumulation allows determining $R_{bulk}$. A value of ~3.7 k$\Omega$ is found, which is about a third of the value obtained from a naïve 1-D model: $R_{bulk} = (\rho_{bulk} \cdot d)/A$; $d$ and $\rho_{bulk}$ are the thickness and the resistivity of the non-depleted bulk, and $A$ the area of the aluminum on top of the isolator. The difference is not yet understood quantitatively. A 3-D simulation shows that the accumulation layer under the $SiO_2$ extends beyond the aluminum contact and thus the effective area relevant for $R_{bulk}$ is significantly larger than $A$. For the calculation of the C/G-V curves the value of $R_{bulk}$ determined from the measured conductance is used. For completeness $C_{bulk}$, the capacitance of the non-depleted silicon bulk, is also included in the model.

From the values of $Q_{it}^j$, $C_{it}^j$ and $G_{it}^j$ obtained from the TDRC measurements as discussed above, the shapes of the C/G-V curves for different frequencies are calculated using the model shown in Fig. 1. Fixed oxide charges only shift the C/G-V curves. Thus the oxide charge $Q_{ox}$ can be determined from the voltage shift between the measured and the calculated curves and the oxide charge density from $N_{ox} = Q_{ox}/A$. As illustration, Fig. 2 compares the measured C/G-V curves for the MOS capacitor irradiated to 10 MGy, after annealing for 10 minutes at 80 °C, to the result of the calculations using the optimized $Q_{ox}$ value. It is seen, that the model of Fig. 1 provides an adequate description of the data. This gives us some confidence, that the extraction of the parameters is reliable.

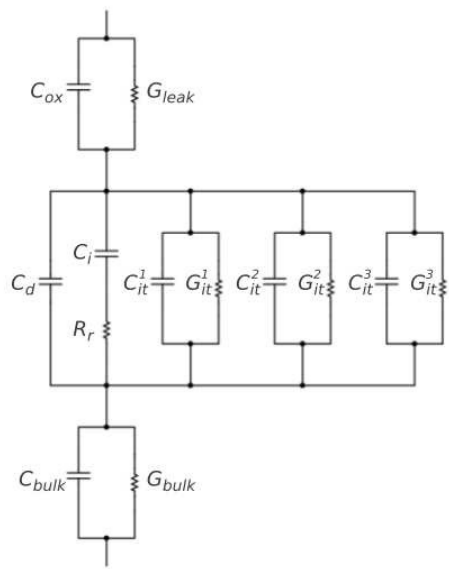

**Fig.1**: Model with three interface traps used to calculate the C/G-V curves for the MOS capacitors. The symbols and the methods used to determine the parameters are described in the text.

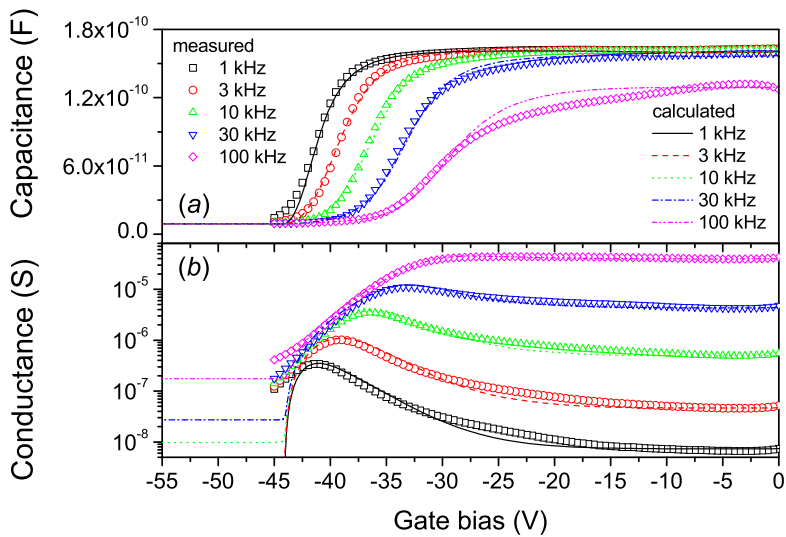

**Fig.2**: Comparison of the measured to the calculated C-V and G-V curves (parallel mode) of a MOS capacitor irradiated to 10 MGy (12 keV X-rays) for frequencies between 1 and 100 kHz using the model displayed in Fig. 1: (*a*) C-V curves, (*b*) G-V curves.

## 3. Results

We first present results from MOS capacitors irradiated to eight X-ray doses in the range between 12 kGy and 1 GGy. The irradiations took between 7.5 seconds and 1.5 hours. The C/G-V measurements were performed about 20 minutes after the irradiations. Due to a significant annealing of the radiation induced effects already at room temperature, one has to be careful with a quantitative interpretation of the data. Fig. 3 shows the C-V and G-V data for frequencies of 1 and 10 kHz. As discussed below, the voltage scans are stopped before strong inversion is reached, to avoid the injection of holes into border traps located in the $SiO_2$ close to $Si$-$SiO_2$ interface (Fleetwood *et al.*, 1994; Fleetwood *et al.*, 2008).

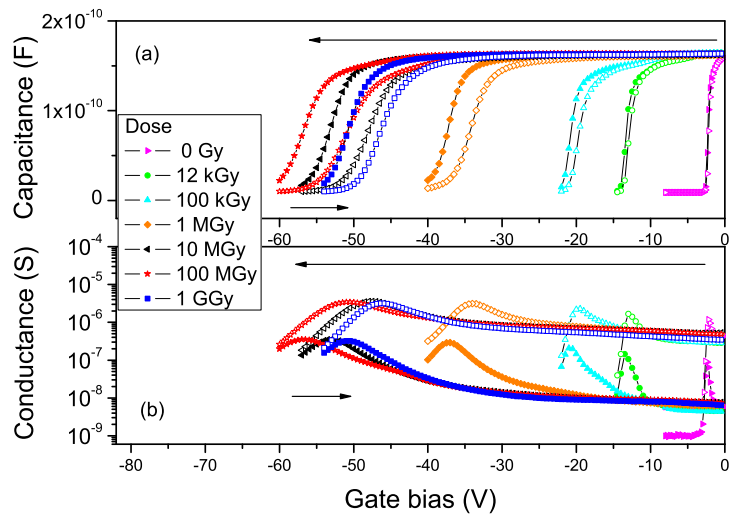

**Fig.3**: (*a*) C-V and (*b*) G-V curves of the MOS capacitors as a function of dose (12 keV X-rays) measured at 1 kHz and 10 kHz (full and open symbols) 20 minutes after the irradiation.

The following qualitative observations can be made: The unirradiated capacitor shows C/G-V curves close to what is expected for an ideal MOS capacitor with an oxide charge density of $1.3 \times 10^{11} cm^{-2}$ (see below). The transition from accumulation to inversion occurs within ~0.5 V and only a small frequency shift due to interface traps is seen. With irradiation the curves shift to negative voltages, the transition becomes shallower and a strong frequency shift is observed. This is a clear evidence for positive oxide charges and interface traps, which can be well described by the model discussed above. The maximum voltage shift occurs for a dose of 100 MGy; the 1 GGy curves are shifted backwards by about 5 V. The reason is probably annealing at high dose rates during the longer irradiation time as discussed below.

For the unirradiated capacitor the following parameters, which characterize the technology, are obtained: The constant capacitance of 163.4±0.3 pF at small negative voltages, where the MOS structure is in accumulation, corresponds to $C_{ox}$, the capacitance of the SiO$_2$-Si$_3$N$_4$ layer. The value found, agrees with the expectation from the area $A$, the SiO$_2$ and Si$_3$N$_4$ thicknesses and dielectric constants. At higher absolute voltages, where the MOS structure is in strong inversion, a value of 8.80±0.02 pF is measured, from which an effective doping of the silicon close to the Si-SiO$_2$ interface of $6.9 \pm 1.4 \times 10^{11}$ cm$^{-3}$ is derived. The value agrees with the bulk doping of $7 \times 10^{11}$ cm$^{-3}$ obtained from the C-V curve of a nearby pad diode. From the doping we calculate a value of $C_{FB}$ = 30.9 pF for the flat-band capacitance. The flat-band voltage, $V_{FB}$, for the MOS capacitance before irradiation is -2.25 V, corresponding to an oxide charge density of $1.3 \times 10^{11}$ cm$^{-2}$. This value is typical for the <100> orientation.

When performing the C/G-V measurements it has been observed, that for irradiated MOS capacitors biased in strong inversion holes are injected into the border traps (Fleetwood *et al.*, 1994; Fleetwood *et al.*, 2008). They cause further shifts of the C/G-V curves. The effect is illustrated by the C-V curves shown in Fig. 4(*a*) for a MOS capacitor irradiated to 5 MGy and annealed at 80 °C for 30 minutes. The gate voltage is cycled four times from 0 V to maximum voltages $V_{max}$ and back to 0 V, with $V_{max}$ values of -40, -60, -80 and -80 V. The shapes of the four curves (and their frequency dependences – not shown) are identical, but the voltages, at which the flat-band capacitance is reached change. They are -36, -36, -40 and -44 V, which is evidence for additional positive oxide charges. The interpretation, that the voltage shifts are due to the injection of holes into border traps when the capacitor is biased in strong inversion, is confirmed in Fig. 4(*b*), where TDRC spectra for different gate voltages are shown. The TDRC signal around 270 K increases with higher negative gate voltages. It has been observed, that it takes about 10 days at room temperature or one hour at 80 °C to discharge these deep traps in irradiated MOS capacitors.

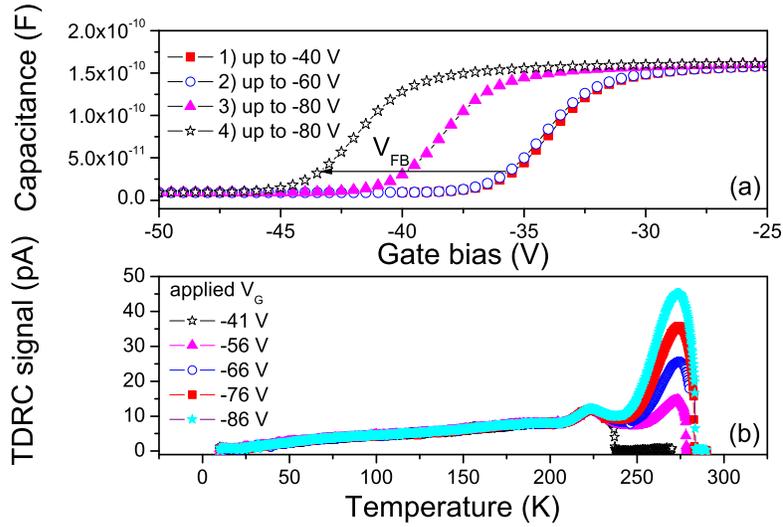

**Fig.4:** (*a*) C-V curves for 1 kHz of the MOS capacitor irradiated to 5 MGy after annealing at 80 °C for 30 minutes biased from 0 V to different maximum voltages. (*b*) Corresponding TDRC spectra. The voltage shifts and the increase of the TDRC signal are evidence for the injection of holes into the border traps, when the capacitor is biased to strong inversion.

In order to obtain reproducible results and a reliable determination of the fixed oxide charge densities $N_{ox}$, the following procedure has been adopted: Annealing of the samples for 10 minutes at 80 °C and stopping the voltage scan before strong inversion is reached. As criterion we use the voltage at which the measured capacitance at 1 and 100 kHz are approximately equal, as seen in Fig. 2. This procedure gives reproducible results, minimizes the effects of border charges and reduces the uncertainties due to the short-term annealing of radiation induced effects.

Fig. 5(*a*) shows the TDRC signal distribution for a MOS capacitor irradiated to 100 MGy and annealed for 10 minutes at 80 °C. Using equations 1 - 3 the $D_{it}(E_t)$ spectrum is extracted. As shown in Fig. 5(*b*), its main features can be described by three states with Gaussian energy distributions $D_{it}^1$, $D_{it}^2$ and $D_{it}^3$ as given in Table 1. There is also a peak around 100 K, which, however as acceptor close to the conduction band, has no influence at room temperature. It should be noted here, that the fit to the TDRC spectrum does not allow an unambiguous determination of the energies and widths of the different trap levels. In addition, the spectra at low temperatures change in a way which cannot be described with a single level. Thus this analysis should not be considered as unambiguous determination of the radiation induced interface traps, but as a consistent, phenomenological description of the measurements.

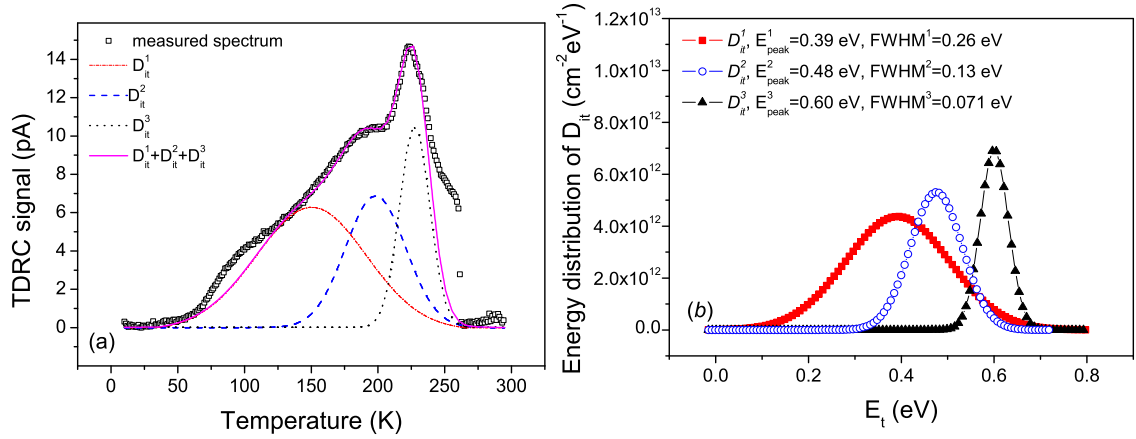

**Fig.5:** (*a*) TDRC spectrum for a MOS capacitor irradiated to 100 MGy with a fit by three dominant interface trap levels. (*b*) The shapes of the three dominant interface trap levels as function of their position in the silicon band gap measured from the conduction band.

For the calculation of the energy distributions and of the densities of traps, the frequency factors in equations 1 - 3 have to be known for every state. They can be determined from the TDRC spectra measured for different heating rates (e.g. $\beta_1$ and $\beta_2$) following the procedure described in Uranwala *et al.* (1975). Thus, the frequency factor for each of the trapping centers is estimated to be:

$$\nu = 10^y \text{ with } y = (T_2 \log_{10}\beta_2 - T_1 \log_{10}\beta_1)/(T_2 - T_1) \tag{9}$$

where $T_2$ and $T_1$ are the TDRC peak temperatures for the heating rates $\beta_1$ and $\beta_2$.

This method is only reliable, if the trap levels do not overlap. However, they do, as seen in Fig. 5. In the annealing study of MOS capacitors fabricated on <111> silicon it has been observed, that after annealing for 32 hours at 80 °C the signal from $D_{it}^2$ is significantly reduced and the signals from $D_{it}^1$ and $D_{it}^3$ are separated (Fig. 6). Thus the frequency factors for these two interface states could be determined (using eq. 9) by performing TDRC measurements with different heating rates $\beta$ (between 0.1 and 0.5 K/s) after annealing for 32 hours at 80 °C. The estimated ranges for the frequency factors are $10^{11.6}$-$10^{12}$ s$^{-1}$ and $10^{11.5}$-$10^{12.5}$ s$^{-1}$ for $D_{it}^1$ and $D_{it}^3$, respectively. These frequency factors are used in eq. 2 to calculate the capture cross section for majority carriers at different temperatures. The values for 295 K, where the C/G-V have been measured, are given in Table 1. In the estimation, the frequency shifts of the C/G-V curves have also been taken into account. With these values, the capture cross section of $D_{it}^2$ has then been obtained by comparing the calculated C/G-V curves to the measured ones for all investigated samples after annealing for 10 minutes at 80 °C. Table 1 summarizes the values estimated for the cross sections, the peak activation energies and the Full Widths at Half Maximum (FWHM). When performing the annealing studies on the <100> crystals it has been noticed, that after annealing for 72 hours at 80 °C a strong $D_{it}^2$ signal remains. Thus the method for determining the cross sections described above cannot be used here and for that reason the <111> values have been used in the analysis.

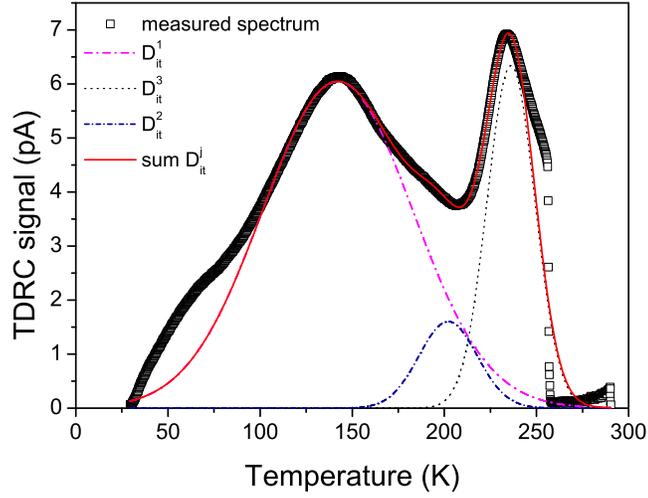

**Fig.6:** TDRC spectrum for a MOS capacitor fabricated on <111> silicon irradiated to 4.8 MGy after annealing for 32 hours at 80 °C: The signal due to the $D_{it}^2$ state is significantly reduced due to annealing.

**Table 1:** Properties of the three dominant interface trap levels.

|  | $D_{it}^1$ | $D_{it}^2$ | $D_{it}^3$ |
|---|---|---|---|
| capture cross section σ [cm²] | 1.2×10⁻¹⁵ | 5×0⁻¹⁷ | 1.0×10⁻¹⁵ |
| peak energy [eV] | 0.39 | 0.48 | 0.60 |
| FWHM [eV] | 0.26 | 0.13 | 0.071 |

Fig. 7 shows the main results of the study: The oxide charge density $N_{ox}$ and the integrated interface trap densities

$$N_{it}^j = \int_0^{E_g} D_{it}^j(E_t)\, dE_t \tag{10}$$

as function of the 12 keV X-ray dose after 10 minutes annealing at 80 °C. It is found that all densities saturate for doses between 10 and 100 MGy with the exception of $N_{it}^2$, which shows a maximum at about 100 MGy and then decreases for the 1 GGy dose. This specific behavior can possibly be explained by annealing during the long-term irradiations at the high dose rate required to achieve high doses. The irradiation may locally heat the sample and cause the annealing of $N_{it}^2$, which is faster than for the other states. A similar study has been performed on circular gated diodes produced on <111> n-type silicon. The same three dominant interface traps with parameters compatible with the ones from the <100> MOS capacitors can describe the data. Also the dependences of $N_{ox}$ and of the $D_{it}^j$ on dose are similar. The values for the <111> material are approximately 20-30 % higher than for <100>.

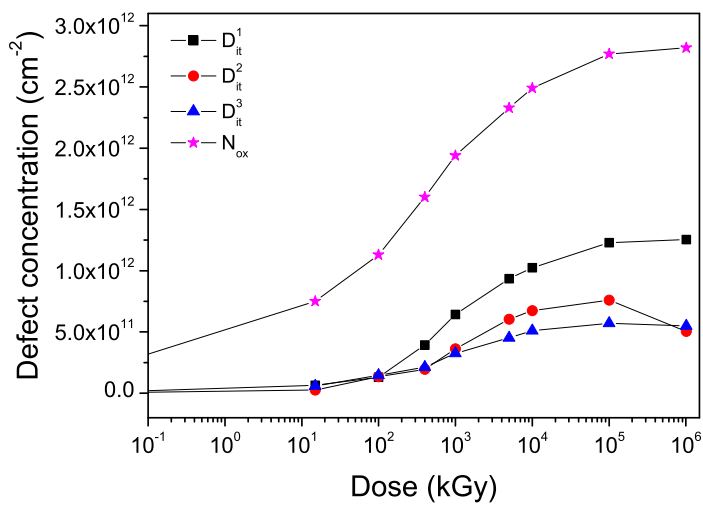

**Fig.7:** Dose dependence of the fixed oxide charge density $N_{ox}$ and the three dominant X-ray induced interface states as function of irradiation dose after annealing at 80 °C for 10 minutes

## 4. Conclusion

Results on the densities of oxide charges and Si-SiO$_2$-interface traps in MOS capacitors built on high-ohmic <100> and <111> n-type silicon as function of the 12 keV X-ray dose up to 1 GGy have been presented. The measurement techniques used are TDRC (Thermal Dielectric Relaxation Current) and C/G-V (Capacitance/Conductance-Voltage) for different frequencies. It is found, that in order to achieve reproducible results, a short annealing step at 80 °C is required, and the capacitors should not be biased to strong inversion to avoid the injection of holes into border states. In addition to radiation induced oxide charges, at least three dominant interface traps have to be used to describe the measurements. Their properties and densities as function of X-ray dose have been determined. The oxide charge density and the trap densities saturate and for one interface state even decrease at dose values between 10 and 100 MGy. The densities of radiation induced traps and oxide charges are approximately 20-30 % higher for the <111> material than for the <100> one. The shape of the C/G-V curves can be described by a simple model with parameters derived mainly from the TDRC measurements. These parameters are presently implemented in a simulation program to check if the performance of segmented sensors as a function of X-ray dose can be described and if the results can be used to design radiation hard silicon sensors.

**Acknowledgements**

This work was done within the Project "Radiation Damage" financed by the XFEL-Company in close collaboration with the AGIPD project (AGIPD webpage). Additional support was provided by the Helmholtz Alliance "Physics at the Terascale". J. Zhang is supported by the Marie Curie Initial Training Network "MC-PAD" (MC-PAD webpage). I. Pintilie gratefully acknowledges the financial support from the Romanian National Authority for Scientific Research through the CNCSIS Project PCCE ID_76.

  We would like to thank G. Potdevin, A. Rothkirch and U. Trunk for their help in setting-up and running the irradiation facility, F. Januschek, F. Renn and T. Theedt for contributions at the early stage of the project and W. Gärtner for the design and the fabrication of the irradiation stand.